# Electromagnetic-radiation absorption of water

P. Lunkenheimer,* S. Emmert, R. Gulich, M. Köhler,[†] M. Wolf,[‡] M. Schwab, and A. Loidl

*Experimental Physics V, Center for Electronic Correlations and Magnetism, University of Augsburg, 86159 Augsburg, Germany*

Why does a microwave oven work? How does biological tissue absorb electromagnetic radiation? Astonishingly, we do not have a definite answer to these simple questions because the microscopic processes governing the absorption of electromagnetic waves by water are largely unclarified. This absorption can be quantified by dielectric loss spectra, which reveal a huge peak at a frequency of the exciting electric field of about 20 GHz and a gradual tailing off towards higher frequencies. The microscopic interpretation of such spectra is highly controversial and various superpositions of relaxation and resonance processes ascribed to single-molecule or molecule-cluster motions have been proposed for their analysis. By combining dielectric, microwave, THz, and far-infrared spectroscopy, here we provide nearly continuous temperature-dependent broadband spectra of water. Moreover, we find that corresponding spectra for aqueous solutions reveal the same features as pure water. However, in contrast to the latter, crystallization in these solutions can be avoided by supercooling. As different spectral contributions tend to disentangle at low temperatures, this enables to deconvolute them when approaching the glass transition under cooling. We find that the overall spectral development, including the 20 GHz feature (employed for microwave heating), closely resembles the behavior known for common supercooled liquids. Thus, water's absorption of electromagnetic waves at room temperature is not unusual but very similar to that of glass-forming liquids at elevated temperatures, deep in the low-viscosity liquid regime, and should be interpreted along similar lines.

## I. INTRODUCTION

Water is important. This simple statement is self-evident but, nevertheless, many of water's unusual physical properties are not well understood. Among these, its dielectric properties play a prominent role. For example, the so-called dielectric loss $\varepsilon''$ (the imaginary part of the dielectric permittivity) quantifies the "loss" of field energy due to absorption by the sample. It is not only relevant for daily-life applications as microwave cooking [1], airport body scanners, or the biological effects of mobile-phone radiation, but also for less-known but nevertheless important effects, as, e.g., the damping of electromagnetic waves by fog or clouds, affecting communication and radar devices [2]. From a more fundamental point of view, it is astonishing that even the microscopic origin of the dominating absorption mode, revealed in the room-temperature dielectric-loss spectra of water at about 20 GHz [Fig. 1(b)], is controversially discussed, not to mention the faster dynamic processes detected in such spectra if extending to sufficiently high frequencies beyond THz.

The 20 GHz feature shows up as a strong peak in the frequency-dependent dielectric loss $\varepsilon''(\nu)$ [Fig. 1(b)]. Its spectral shape can be relatively well fitted by a Debye relaxation-function [3] and it is often ascribed to the so-called $\alpha$ or structural relaxation, reflecting the molecular dynamics that governs, e.g., the viscosity of a liquid [4,5]. (See, e.g., Ref. 6 for a comprehensive data collection on its temperature dependence and for its relevance for the controversially discussed glass temperature of water.) However, very recently earlier ideas [1,7,8] were revived [9] explaining this spectral feature in a completely different way, namely along similar lines as the Debye relaxation known to arise in most monohydroxy alcohols [10,11,12]. This relaxation process is ascribed to the dynamics of clusters formed by hydrogen-bonded molecules, which is significantly slower than the $\alpha$ relaxation, essentially arising from single-molecule motions [13]. Interestingly, the 20 GHz relaxation in water does not, or only weakly show up in susceptibility spectra obtained by light-scattering or optical-Kerr-effect measurements [9,14,15,16,17,18], similar to the findings in the alcohols.

Moreover, as suspected since long [19] and recently clearly revealed by broadband spectra extending into the THz range [8,15,18,20,21,22,23,24], at the high-frequency flank of the dominating 20 GHz loss peak (between about 300 GHz and 2 THz), excess intensity is detected, indicating contributions from a faster dynamic process. It shows up as a second, more shallow power law, superimposed to the $\nu^{-1}$ decrease found for a Debye relaxation [Fig. 1(b)] and will be termed high-frequency power law (HFPL) in the following. Within the scenario described above, assuming a cluster-related origin of the 20 GHz relaxation, this faster process could represent the "true" $\alpha$ relaxation, i.e., the dynamics of single water molecules [1,7,9]. Indeed, the data up to about 1 THz can be fitted by the sum of two Debye functions, where the second relaxation is strongly superimposed by the dominating 20 GHz process [2,4,5,8,22,25,26,27]. (At even higher frequencies, several vibration-related processes are detected [2,14,19,20,21] which can be fitted by Lorentzian functions but are not within the scope of the present work.) In contrast, in light-scattering or optical-Kerr-effect spectra, being much less sensitive to the

---
*Corresponding author.
peter.lunkenheimer@physik.uni-augsburg.de
[†]Present address: Osram GmbH, 86153 Augsburg, Germany
[‡]Present address: Instrument Systems GmbH, 81637 Munich, Germany

20 GHz feature, this faster relaxation is clearly revealed by a well-pronounced shoulder [9,15,17,18]. For this process, relaxation rates (i.e., loss-peak frequencies) between 150 and 940 GHz were deduced from dielectric spectroscopy [2,4,8,15,20,21,22,25,26,27] while the other methods yield rates around 200 - 400 GHz [9,15,17]. Notably, in some publications [15,20,21,28], in addition to a second Debye process, the HFPL in the dielectric loss of water was analyzed assuming even another fast process, which is located at about 1.3 - 1.9 THz.

It was recently pointed out [23] that the HFPL in dielectric spectra of water closely resembles the so-called excess wing (EW) found in dielectric loss spectra of many supercooled liquids [29,30,31]. This phenomenon is often assumed [32,33,34,35,36,37] to arise from a so-called $\beta$ relaxation [38], partly submerged under the dominating $\alpha$ peak. Numerous further contradicting interpretations of the main 20 GHz relaxation, the suggested second faster relaxation, and the possible 1.3 - 1.9 THz excitation were proposed in literature [5,15,18,20,22,23,27,28,39]. Except for the dominating 20 GHz feature, even the mere existence and the spectral form of the two latter processes is still controversial: In fits of broadband spectra, the HFPL was accounted for by using both processes simultaneously [9,15,20,21,28], only the fast Debye [2,22], or only the 1.3 - 1.9 THz excitation [18]. Moreover, for the latter, either a relaxational [20,21] or a resonance character [9,15,18,28] were assumed. It should be noted that a process with a rate of 1.1 - 1.8 THz in water was also deduced by molecular-dynamics simulations, neutron, and light scattering, whose interpretation, however, is also controversial [16,17,28,40,41,42,43,44,45].

Obviously, there is much confusion concerning the microscopic dynamic processes governing the interaction of water with electromagnetic waves in this technically relevant frequency region, a highly dissatisfactory situation. In the present work, we tackle this problem in a twofold way: At first, by combining dielectric, microwave, THz, and far-infrared techniques, we provide nearly continuous dielectric spectra for different temperatures down to the freezing point, covering the main relaxation, the HFPL, and the two first clearly visible infrared resonances. The mentioned problems in the interpretation of dielectric broadband spectra of water partly arise from the fact that these are usually composed of results from different sources, often not taken at the same temperature. This leads to severe ambiguities in the construction of continuous spectra and their modelling, which is not the case for our data. In addition, nearly all broadband spectra reported in literature are limited to room temperature. Temperature-dependent data are only available in restricted frequency ranges, either missing the 20 GHz peak or the resonances beyond THz [4,5,20,22,23], which, however, both are important for a correct analysis of the spectra.

Moreover, and most importantly, we have also performed corresponding broadband experiments extending beyond THz for water mixed with LiCl, which we find to reveal the same general spectral behavior as pure water. However, in contrast to pure water, where crystallization under cooling leads to the so-called "no-man's land" [46] where the supercooled-liquid state cannot be investigated, such aqueous salt solutions can be easily supercooled down to low temperatures approaching the glass state [47,48]. Supercooling leads to a shift of the 20 GHz relaxation peak and any other possible relaxation processes to arbitrarily low frequencies. This should enable a better identification of the suggested processes at 150 - 940 GHz and/or 1.3 - 1.9 THz because faster processes usually exhibit weaker temperature dependence than the main relaxation [49]. The same can be said for resonance processes compared to relaxational ones, the latter being essentially thermally activated and exhibiting much stronger temperature dependence. For the monohydroxy alcohols, the $\alpha$ relaxation, which at high temperatures is strongly superimposed by the Debye peak, usually also is more clearly disclosed at low temperatures (see, e.g., refs. [11,12]). Our experiments reveal that the HFPL is due to an absorption process, termed boson peak, known from other dipolar liquids, and that, generally, the room-temperature absorption spectra of water should be explained in a similar way as the spectra of supercooled liquids at elevated temperatures.

## II. EXPERIMENTAL DETAILS

Aqueous 8 mol/l LiCl solution was purchased from Sigma Aldrich. For dilution and for the measurements of pure water, deionized $H_2O$ (Merck "Ultrapur") was used.

To obtain broadband dielectric spectra, several techniques were combined: At the lowest frequencies (< 1 MHz) frequency-response analysis (Novocontrol "Alpha-A" analyzer) was used. The interval up to 3 GHz was covered by a coaxial reflectometric setup [50,51] with the two impedance analyzers Agilent 4294A and Agilent E4991A. For these two methods the samples were prepared as parallel-plate capacitors. The plates were made of stainless steel and had typical diameters of 0.4 - 2 cm and plate distances of 100 - 500 µm. The temperature was controlled via a $N_2$-gas flow cryostat.

The microwave frequency range (100 MHz - 40 GHz) was covered by the Agilent "Dielectric Probe Kit" in combination with the Agilent E8363B Network Analyzer. Here the reflection coefficient of an open-ended coaxial line, whose end is directly immersed into the liquid sample, is measured. Within this work, the so-called "Performance Probe" was used. The dielectric quantities $\varepsilon'$ and $\varepsilon''$ were calculated taking into account the electric-field distribution at the end of the line [52,53]. The temperature was varied via a Peltier element.

At higher frequencies, a quasi-optical spectrometer developed by Volkov and coworkers [54] was used. It follows a Mach-Zehnder setup, which allows for the independent determination of the transmission coefficient and phase shift of the radiation passing the sample. So-called "backward wave oscillators" (BWOs) are used as sources. At the open end of the waveguides integrated in the BWOs, the radiation penetrates into free space. The radiation frequency can be changed by the variation of the voltage but the bandwidth is limited by the waveguide. Six different BWOs were used to cover the frequency range 50 GHz $\leq \nu \leq$ 1.2 THz. The signal was detected by a pumped He bolometer. Temperature variation was achieved by a home-made oven and nitrogen cryostat.

The terahertz time-domain spectrometer TPS Spectra 3000 by Teraview Ltd. covers the spectral interval from 100 GHz to 3 THz. A femtosecond laser radiated at a GaAs substrate produces "white" THz pulses, which are detected with variable time delay. Fourier transformation of the time-domain signal leads to the frequency-dependent intensity and phase angle. For temperature-dependent measurements below room temperature, a helium-flow cryostat was employed. To reach higher temperatures, a heating coil was placed into the sample compartment.

For the measurements with the Mach-Zehnder setup and the Teraview spectrometer, the liquid samples were put into sample cells with thin quartz windows. Due to the strongly varying transmission, depending on frequency and temperature, the thickness of the used sample cells had to be adapted between 0.2 and 1 mm. The results were corrected for the contributions of the windows (including multiple reflections), whose properties can be determined by measurements of the empty cuvettes and/or the separate windows. For this three-layer system, the dielectric properties of the samples were extracted via standard optical formulae for multilayer interference. For this purpose, the program "ThreeLayer" by Y. Goncharov was used.

For the measurements of pure water above 1 THz, the Fourier-transform infrared spectrometer Bruker IFS 113v was employed. By using a heated silicon carbide rod as a black body source and a liquid-helium cooled Si-bolometer as detector, the frequency range up to 20 THz can be covered. For the temperature-dependent transmission measurements a nitrogen-flow cryostat, equipped with a pair of 1 mm thick high-density polyethylene windows, was used. At these high frequencies, the sample thickness has to be as thin as possible to ensure sufficient transmission despite the high absorption of water in this region. Thus, the sample was prepared as a 14 µm layer sealed within a home-designed cuvette made of polyethylene. By knowing the temperature-dependent optical properties of the cuvette material, the transmittance of the water layer can be deduced from the spectrum of the filled cuvette. In contrast to the quasi-optical technique and terahertz time-domain spectroscopy, with the Fourier-transform infrared spectrometer it is only possible to measure the transmittance of the sample, while the phase shift cannot be determined. For this reason, a Kramers-Kronig transformation was applied leading to the complex refractive index, from which the complex dielectric constant was calculated. For the Kramers-Kronig transformation, a proper choice of low- and high-frequency extrapolations is essential. The low-frequency extrapolation was based on the experimental data of the present work, while at high frequencies the data were extrapolated to match the known index of refraction at optical frequencies.

## III. RESULTS

### A. Pure water spectra

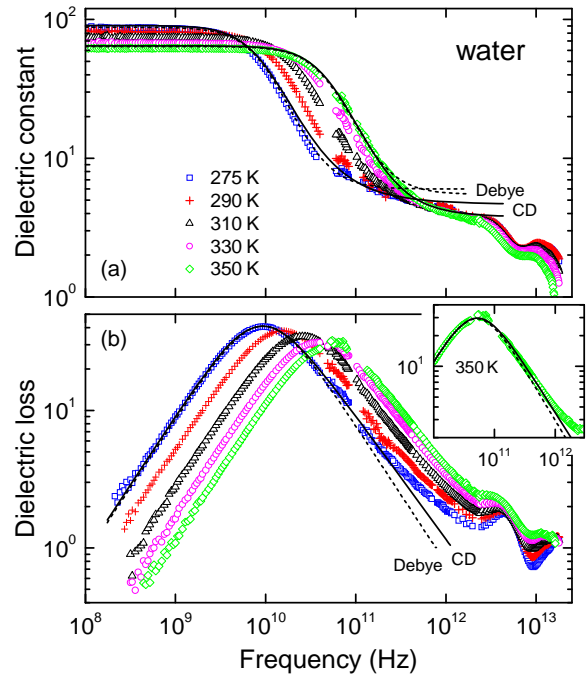

FIG. 1. Dielectric permittivity of pure water. (a) Dielectric constant and (b) loss spectra of water at selected temperatures (for more temperatures, see Fig. 5 in Appendix A). The dashed and solid lines are fits with the Debye and CD functions, respectively. Inset: Loss for 350 K with fits.

Figure 1 shows the dielectric constant $\varepsilon'(\nu)$ and loss $\varepsilon''(\nu)$ of pure water at selected temperatures from 275 to 350 K (see Fig. 5 for all investigated temperatures). Results from all available experimental methods were combined to obtain these spectra extending from 100 MHz to 20 THz. We want to point out that, in contrast to most other published broadband water spectra, these data stand out by being measured by a single workgroup only, not including any results from literature. Moreover, results for various temperatures are provided, which are identical for all experimental frequency ranges, and ambiguities arising from the application of the Kramers-Kronig relation in the infrared range are minimized by the availability of independently measured real- and imaginary-part data at lower frequencies. In Fig. 1, the main relaxation process and its slowing down with decreasing temperature as found in many other dipolar liquids [11,12,29,30,49] are clearly revealed by the dominating step-like decrease in $\varepsilon'(\nu)$ and peak in $\varepsilon''(\nu)$, both shifting to lower frequencies. As examples, for 275 and 350 K the dashed lines show fits by the Debye function, simultaneously performed for $\varepsilon'(\nu)$ and $\varepsilon''(\nu)$. This leads to a good description of the data at low frequencies and up to about one decade above the peak frequency. However, at higher frequencies the mentioned HFPL shows up in the loss, accompanied by an additional decrease of $\varepsilon'(\nu)$, not covered by the fits. Finally, around 5 and 19 THz the first two of the well-known vibration-related high-frequency resonances [2,15,20,21,22] are detected.

These data also demonstrate that the notion of a pure Debye character of the 20 GHz relaxation is somewhat misleading as the spectra can be better fitted for about one additional frequency decade beyond the peak by the empirical Cole-Davidson (CD) function often used for dipolar liquids [29,49,55]. The obtained width parameters are $\beta = 0.90$ (275 K) and $\beta = 0.95$ (350 K) signifying small but significant deviations from $\beta = 1$ expected for the Debye function [55]. A similar finding was reported in Ref. 56. Nevertheless, even for this approach the HFPL still is well revealed as excess intensity in the loss around 1 THz. As mentioned above, in literature broadband spectra of water were fitted by different combinations of relaxation and resonance functions [2,9,15,18,20,21,22,27], with only marginal differences in the agreement of fit and experimental data. 275 K is the lowest temperature covered by the present investigation, where all processes should be maximally separated. Thus, for pure water the fitting of this data set represents the most rigorous benchmark for testing models for the explanation of its dielectric response (a fit example for the present data is shown in Fig. 6 in Appendix B). Notably, the separation of processes is much less pronounced for room-temperature spectra, which were analyzed in most previous works. However, even for 275 K it is clear that, due to the still substantial superposition of the assumed processes beyond the main relaxation, broadband data of water alone do not allow for unequivocal conclusions on the nature of the HFPL of water.

Finally, we want to point out the existence of an isosbestic point in the frequency dependence of the dielectric constant close to 550 GHz for the given temperature range. There $\varepsilon'$ is independent of temperature from the freezing point up to 350 K. For all measured temperatures, this is even more convincingly documented in Fig. 5 (Appendix A). This finding opens up interesting perspectives for scaling approaches as pointed out in Ref. 57.

## B. Concentration-dependent microwave and THz spectra of aqueous LiCl solutions

Following the mentioned approach of adding LiCl to water to resolve the different contributions to its spectra, in a first step Fig. 2 demonstrates that the relevant spectral features indeed also show up in these solutions. It presents room-temperature data for different salt contents, combining results from coaxial reflectometry and time-domain THz spectroscopy. The spectra indeed reveal the presence of the main relaxation and of the HFPL. For high salt concentrations, the latter is even more pronounced than in pure water, demonstrating that this system is ideally suited to investigate its origin.

While the relaxation rate in Fig. 2 is only weakly (less than a factor of 2) affected by ion addition, the amplitude of the 20 GHz relaxation clearly diminishes with increasing salt concentration, a well-known finding for aqueous solutions [24,58,59], which can be ascribed to hydration and depolarization effects [58]. In contrast, the loss at frequencies beyond about 1 THz changes only insignificantly when LiCl is added. This indicates that the proposed excitation feature at about 1.3 - 1.9 THz [15,20,21,28] remains almost unaffected by increasing ion concentration. Interestingly, in the glass-forming dipolar liquid glycerol, LiCl addition also causes a reduction of the main relaxation amplitude (the $\alpha$ relaxation) [60] while the loss above 1 THz, in the regime of the so-called boson peak, varies only weakly [61].

Again, in the dielectric constant an isosbestic point appears close to 60 GHz, pointing towards a complete independence of $\varepsilon'$ as function of molar LiCl concentration.

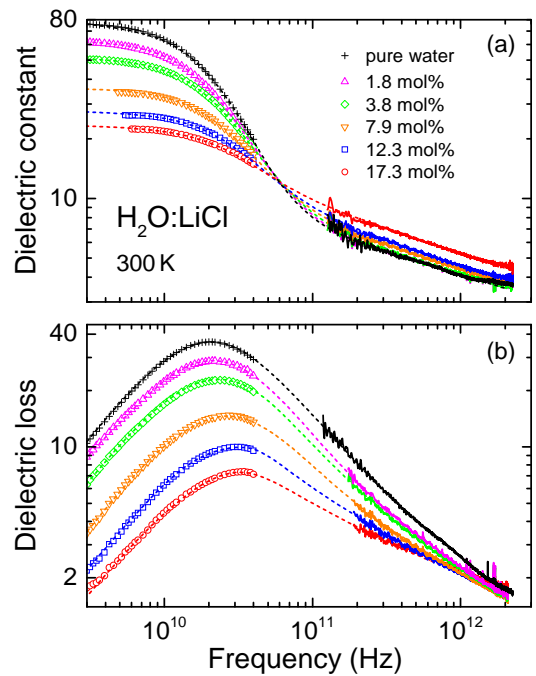

FIG. 2. Concentration-dependent dielectric spectra of $H_2O$:LiCl solutions. (a) Dielectric constant and (b) loss spectra of pure water and water mixed with various amounts of LiCl at room temperature. The shown loss was corrected for the charge-transport contribution by subtracting $\sigma_{dc}/\omega\varepsilon_0$ (with $\sigma_{dc}$ the dc conductivity, $\omega$ the circular frequency, and $\varepsilon_0$ the permittivity of vacuum). Symbols: results from coaxial reflectometry; solid lines: time-domain THz spectroscopy. The dashed lines are guides for the eyes.

## C. Broadband spectra of an aqueous LiCl solution

Figure 3(a) shows broadband loss spectra for a water solution with 17.3 mol% LiCl measured at various temperatures. These results are consistent with those for a 13.7 mol% solution reported in Ref. 48 for frequencies below 2 GHz and $T < 213$ K. The 20 GHz relaxation slows down by more than 12 decades when supercooling this solution. This is indeed expected in both interpretations ($\alpha$ or Debye relaxation) of this spectral feature. For supercooled pure water, relaxation-time ($\tau$) data are available in literature down to about 250 K [6,62]. They are of similar magnitude as $\tau(250\,\mathrm{K}) \approx 33$ ps deduced from the loss-peak frequency $\nu_p$ in Fig. 3(a) via $\tau = 1/(2\pi\nu_p)$. This demonstrates that, also below room temperature (cf. Fig. 2), the dynamics of water is only weakly affected by salt addition. Interestingly, the main loss peak of $H_2O$:LiCl strongly broadens when cooling, especially at its right flank [Fig. 3(a)]. Fits of the main relaxation with the empirical Havriliak-Negami function [63] [black solid lines in Fig. 3(a)] reveal power-law exponents between -0.54 (250 K) and -0.38 (135 K) for the high-frequency flank, clearly deviating from the $\nu^{-1}$ power law expected for the Debye function.

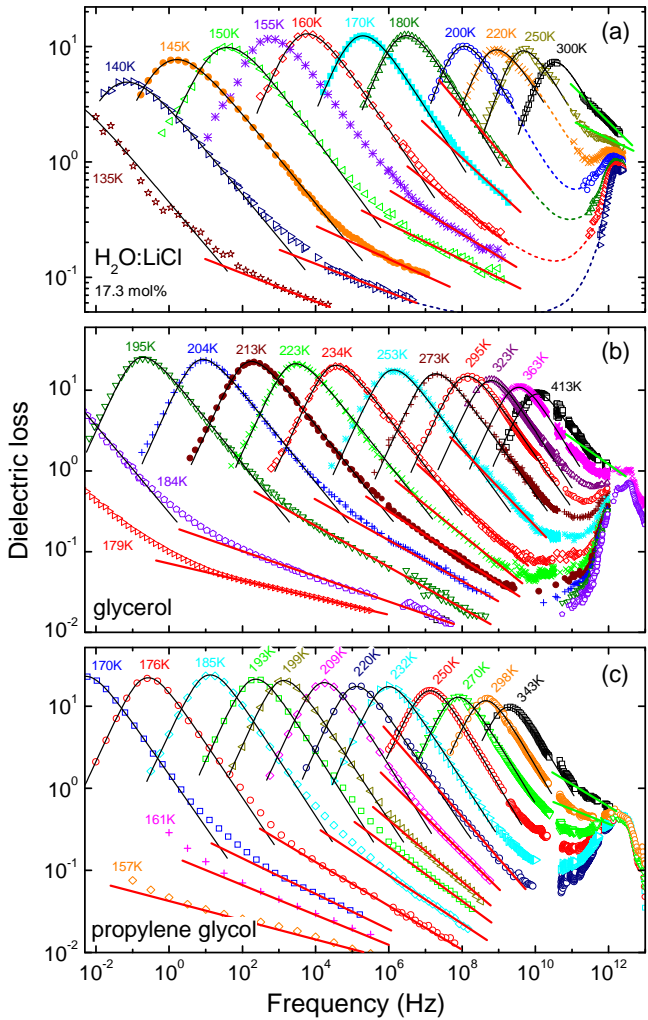

FIG. 3. Broadband dielectric loss spectra. (a) Spectra of a 17.3 mol% water:LiCl solution at various temperatures. The shown loss was corrected for the charge-transport contribution by subtracting $\sigma_{dc}/\omega\varepsilon_0$. The black solid lines are fits with the Havriliak-Negami function. The red solid lines indicate the EW; the green solid lines indicate the HFPL. The dashed lines are guides for the eyes. Frames (b) and (c) show broadband loss spectra of glycerol [30] and propylene glycol [49], respectively.

## IV. DISCUSSION

When assuming that the main relaxation in Fig. 3(a) indeed is cluster related as in monohydroxy-alcohols, in this $H_2O$:LiCl solution it obviously behaves different than in the alcohols, where its Debye spectral form is retained down to low temperatures [11,12]. One may speculate that LiCl addition leads to a distribution of cluster relaxation times and, thus, a broadening of the Debye relaxation. It should be noted, however, that in canonical supercooled liquids, in the low-viscosity liquid state, where the $\alpha$ peak is located in the frequency region beyond GHz, its spectral form also often approaches that of a Debye relaxation and it considerably broadens under cooling only [29,49,64], consistent with the behavior documented in Fig. 3(a).

In monohydroxy alcohols, just as the Debye relaxation, the $\alpha$ relaxation also strongly shifts to low frequencies under cooling [10,11,12]. However, in most cases its temperature dependence does not completely parallel that of the Debye relaxation time and, especially at lower temperatures, the two time scales separate more strongly than at high temperatures. Thus, there the $\alpha$ relaxation often is clearly revealed as a distinct peak or shoulder instead of a mere power law at higher temperatures [11,12]. However, we note that in Fig. 3(a) no such separation of both dynamics occurs. Instead, at low temperatures a second power law at the right flanks of the loss peaks is detected (red solid lines). Within the alcohol scenario, it may be ascribed to the $\alpha$ relaxation, whose low-frequency part is still completely submerged under the cluster relaxation. The situation then would be similar to that in 2-ethyl-1-hexanol, where the $\alpha$-peak amplitude is about two decades lower than the Debye peak [11]. (However, there the $\alpha$ relaxation is revealed as a clear shoulder at low temperatures instead of a mere power law.)

On the other hand, the overall spectral shape and temperature development observed in Fig. 3(a) closely resembles that of typical glass-forming liquids [see Figs. 3(b) and (c) for two examples] [29,30,31,37,49,64]. Such broadband spectra of supercooled liquids are known to exhibit a sequence of dynamic processes, namely $\alpha$ relaxation, $\beta$ relaxation or EW, fast $\beta$ process, and the boson peak [29,30]. Within this scenario, the main peak in the spectra of Fig. 3(a) corresponds to the $\alpha$ relaxation and the mentioned second power law, also found in many glass formers at low temperatures [red solid lines in Figs. 3(b) and (c)], is the EW [29,32,65], already briefly discussed above.

Could this EW be the same phenomenon as the HFPL observed in the LiCl solution and in pure water at 300 K and beyond 100 GHz? We note from Fig. 3(a) that the EW, found at low temperatures, becomes successively steeper with increasing temperature just as in other supercooled liquids [cf. Figs. 3(b) and (c)]. Finally, it seems to merge with the right flank of the $\alpha$ (or cluster-related) relaxation peak somewhat above 180 K. Notably, in Ref. 48 the EW of a 13.7 mol% LiCl solution, detected by light-scattering at low temperatures, was also reported to vanish already far below room temperature. In fact, in broadband measurements of various dipolar liquids [29,49,61,65,66,67], the EW also never shows up at temperatures for which the main relaxation is located in the GHz range [see also Figs. 3(b) and (c)]. Thus we conclude that the HFPL in water cannot be due to the EW. (It should be noted that the above considerations are also valid if interpreting the low-temperature second power law in Fig. 3(a) as manifestation of the $\alpha$ relaxation within the monohydroxy-alcohol scenario: Its spectral contribution becomes indiscernible at high temperatures and cannot be responsible for the HFPL.)

In fact, the HFPL is not a special property of water: The green solid lines in Figs. 3(b) and (c) indicate, for the two typical examples glycerol and propylene glycol, that similar high-frequency excess contributions also show up at the right flanks of the main relaxation peaks in other dipolar liquids, whenever the temperature is high enough to yield an $\alpha$-peak position beyond GHz (see refs. 49,61,65,66,67,68 for further examples). A hint to the nature of this feature is given by the data at lower temperatures [see, e.g., Figs. 3(b) and (c)]: There supercooled liquids universally exhibit an excitation peak located somewhat above 1 THz [29,49,61,65,67,69] that is narrower than for a relaxation but broader than a typical phononic or intramolecular resonance and that is only weakly varying with temperature (the main temperature effect arises from the overlap with the strongly temperature-dependent main relaxation peak). In the dielectric-spectroscopy literature, this feature is often termed boson peak [29,49,61,65,70,71,72], in analogy to the corresponding excitation known from light and neutron scattering [73,74]. Various, partly contradicting explanations for the occurrence of the boson peak were proposed (e.g., refs. [70,75,76,77,78]), however, all assuming a phonon- or vibration-related origin. As revealed by Figs. 3(b) and (c), the HFPL in the supercooled liquids can be clearly ascribed to the boson peak, strongly superimposed by the $\alpha$ peak at high temperatures. (The fast $\beta$ relaxation, leading to the shallow minimum observed at low temperatures [29,30], is fully submerged under the $\alpha$ relaxation at the highest temperatures.)

Is this boson-peak scenario also valid for the HFPL in the $H_2O$:LiCl solution? Low-temperature measurements in the relevant frequency range around THz should help solving this question: Just as for the supercooled liquids, upon cooling all possible, more-or-less thermally activated relaxational contributions to the spectra of Fig. 3(a) ($\alpha$, second Debye, cluster relaxation, or EW) should shift to low frequencies and possible other contributions to the HFPL should become visible. This is indeed confirmed by the low-temperature data from THz spectroscopy ($\nu > 200$ GHz), included in Fig. 3(a): With decreasing temperature, the HFPL develops into a relatively narrow peak (compared to a typical relaxation) at about 1.5 - 2 THz. Obviously, this peak significantly contributes to the HFPL. As mentioned above, an excitation at 1.1 - 1.8 THz in water was also detected by various other methods, for which different explanations were proposed [16,17,40,42,44,45].

As demonstrated by Fig. 4 the occurrence of this peak at low temperatures is not limited to the solution with the highest investigated LiCl concentration of 17.3 mol%. While we have chosen this rather high concentration for the broadband measurements due to its low crystallization tendency, in the THz range, by careful measurements under highly clean conditions, we also succeeded in collecting data in the supercooled state for the lower concentration of 7.9 mol%. The resulting loss spectra are shown in Fig. 4 (triangles). Under cooling, they exhibit the same development from a HFPL to a peak as found for the 17.3 mol% solution (circles). Obviously, the occurrence of this peak does not depend on the salt concentration and it is reasonable to assume that, also in pure water, it should contribute to the observed HFPL. The different slopes of the HFPLs at 300 K, revealed for the two concentrations in Fig. 4, arises from the different amplitudes of the 20 GHz peaks (cf. Fig. 2), corroborating the notion that the HFPL is caused by a superposition of this relaxation peak and the narrow peak at about 1.5 - 2 THz, which is not affected by the salt concentration. At the lowest temperature of 180 K, where the main relaxation feature is shifted to low frequencies and no longer influences the spectra in the THz range [cf. the 180 K curve in Fig. 3(a)], the detected peaks for 7.9 and 17.3 mol% nearly coincide.

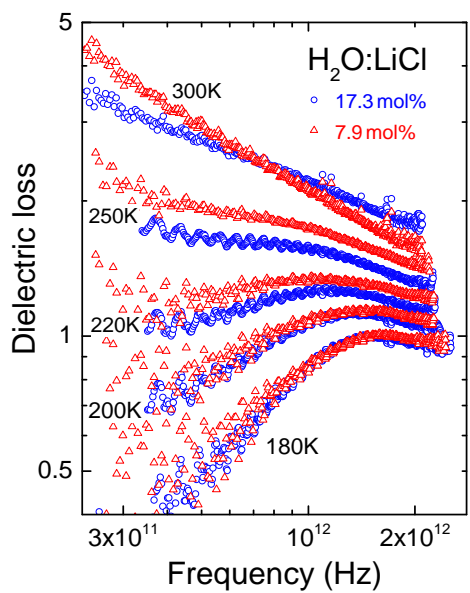

FIG. 4. Temperature dependence of the dielectric loss spectra in the THz region for $H_2O$:LiCl solutions with two different LiCl concentrations.

To make clear that the obtained results on solutions indeed are relevant for the interpretation of pure-water spectra, the following findings should be considered: i) The dominating 20 GHz peak shows up at nearly unchanged frequency in all the solutions, just as in pure water (Fig. 2). ii) The HFPL exists for all concentrations, just as for pure water (Fig. 2). iii) Under cooling, the HFPL develops into a peak for two significantly different concentrations and this peak is not affected by the concentration (Fig. 4). Taking all this together, the analogy between the LiCl solutions and pure water is highly suggestive and it indeed seems reasonable to assume that in pure water the HFPL is also caused by an underlying THz peak, superimposing with the 20 GHz relaxation peak.

The close qualitative similarity of the water-solution spectra in Fig. 3(a) to those of other dipolar liquids [e.g., Figs. 3(b) and (c)] strongly suggests that the peak at 1.5 - 2 THz, contributing to the HFPL in the spectra of the investigated water solutions (and, thus, of pure water), is the same phenomenon as the boson peak, universally observed in other dipolar liquids. The mentioned excitation at 1.3 - 1.9 THz invoked in some publications to fit dielectric spectra of water [15,18,20,21,28] can be identified with this phenomenon. This notion also is in good accord with the interpretation of a corresponding excitation observed by neutron scattering in confined water [42,44,79,80]. The EW does not play a role for the HFPL. Any possible additional processes (second Debye relaxation or $\alpha$ relaxation within the alcohol scenario) only seem to be of limited importance because, even at low temperatures, they do not lead to a separate spectral signature. (However, small deviations of fits of the 275 K spectrum of pure water may indeed indicate some minor additional contribution, see Fig. 6 in Appendix B.) The 20 GHz relaxation peak almost buries the boson peak and only these two processes and the two Lorentzian-type resonances mentioned above govern the dielectric loss up to frequencies of 20 THz.

## V. SUMMARY AND CONCLUSIONS

We have determined the complex dielectric permittivity of pure water and supercooled aqueous solutions in an exceptionally broad frequency range at different temperatures with unprecedented precision. We conclude that there is a boson peak contributing to the dielectric spectra of water and that it plays an important role for the occurrence of the HFPL in the low-viscosity liquid state, just as in other dipolar liquids. Overall, the room-temperature dielectric spectra of water do not seem exceptional when compared to those of common dipolar liquids at elevated temperatures and they should be interpreted along similar lines.

This implies that, at frequencies up to about 100 GHz, the absorption of electromagnetic radiation by water is dominated by the main, nearly Debye-type relaxation process, caused by the overall reorientational motions of the molecular units. At higher frequencies, the boson peak leads to an increasingly important contribution to the dielectric loss (cf. Fig. 6). Here, phonon-like excitations cause the additional absorption. Finally, above about 3 THz, vibrational resonance effects start to dominate the absorption, ascribed to hydrogen-bond stretching (around 5 THz) [22,40] and librational motions, i.e. small-angle molecular rotations (around 20 THz) [22,81,82].

## ACKNOWLEDGMENTS

We thank Ralph Chamberlin for helpful discussions. We gratefully acknowledge the help of Yurii Goncharov in the Mach-Zehnder measurements.

## APPENDIX A: SPECTRA OF PURE WATER AT ADDITIONAL TEMPERATURES

Figure 5 shows dielectric data of pure water as in Fig. 1, however, also including additional temperatures omitted there for clarity reasons.

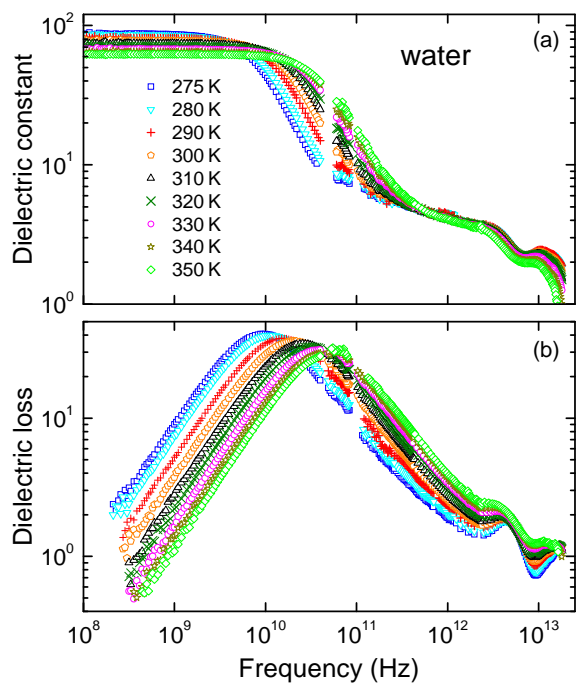

FIG. 5. Dielectric permittivity spectra of pure water at more temperatures. Dielectric constant (a) and loss spectra (b) of water at all investigated temperatures.

## APPENDIX B: FITS OF PURE WATER SPECTRA WITH BOSON-PEAK CONTRIBUTION

The solid lines in Fig. 6 show fits of the water spectra at 275 and 310 K assuming contributions from the main relaxation, two resonances, and the boson peak. The main relaxation was modeled by the Cole-Davidson function, leading to width parameters of 0.87 (275 K) and 0.93 (310 K). For 275 K, it is indicated by the dash-dotted line. The two high-frequency resonances were fitted by Lorentz functions, indicated for 275 K by the dotted lines. For a purely phenomenological description of the boson peak, we used a log-normal peak function with a half width of about 0.9 decades (dashed line), which we found to provide good fits of the boson peaks in dipolar liquids as glycerol or propylene glycol [30,49]. While the fit is nearly perfect for 310 K, at 275 K slight deviations of fit and experimental data show up around 0.3 - 1 THz, which may indicate a contribution from an additional process (e.g., the $\alpha$ process in the monohydroxy-alcohol scenario or a second Debye process). However, one should be aware that the experimental errors are of the order of the symbol size and thus the evidence for this additional process is limited. From such fits, the relative contributions of the different dynamic processes to the loss (and thus to the absorption) can be deduced. Based on the present fits, we estimate for example that at 310 K and 2 THz, 25 % of the absorption arises from the boson peak while the first resonance caused by hydrogen-bond stretching contributes about 14 %.

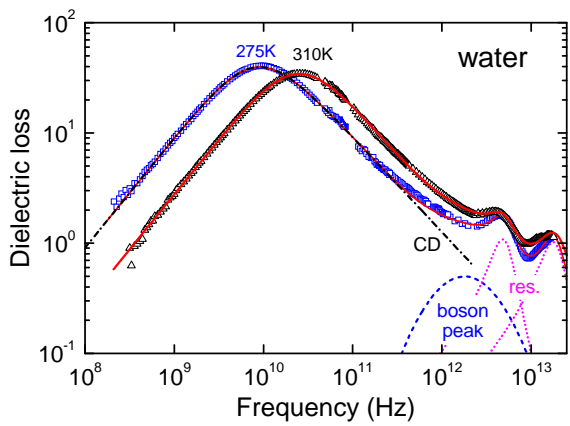

FIG. 6. Dielectric-loss spectra of pure water at 275 and 310 K. The solid lines are fits assuming a CD function (275 K: dash-dotted line) for the main peak and two Lorentz functions (275 K: dotted lines) for the resonances observed at about 5 and 19 THz. In addition, a boson-peak contribution at about 1.8 THz was assumed (dashed line, used for both temperatures).

We want to point out that these fits are only intended to demonstrate the general compatibility of the experimental data with the proposed boson-peak scenario. As discussed in the main paper, due to the substantial superposition of the assumed processes beyond the main relaxation, broadband data of water alone do not allow for unequivocal conclusions on the nature of the high-frequency power law of water.